\def\doctitle{An Open Source C++ Implementation of Multi-Threaded Gaussian~Mixture~Models, k-Means and Expectation Maximisation}
\def\docauthor{Conrad Sanderson, Ryan Curtin}

\documentclass[10pt,a4paper]{article}
\usepackage[includefoot,includehead,a4paper,top=0.5cm,bottom=1cm,left=2cm,right=2cm]{geometry}
\usepackage{fancyhdr}
\usepackage[usenames,dvipsnames]{color}  
\usepackage{titlesec}
\usepackage[latin1]{inputenc}
\usepackage[pdfborder={0 0 0},colorlinks=true,urlcolor=ForestGreen,linkcolor=ForestGreen,citecolor=ForestGreen,bookmarks=true,pdftitle={\doctitle},pdfauthor={\docauthor}]{hyperref}
\usepackage[labelfont=bf]{caption}
\usepackage{enumerate}

\usepackage{graphicx}
\usepackage{amsmath}
\usepackage{amssymb}
\usepackage{url}
\usepackage{mathtools}
\usepackage{xspace}
\usepackage{multirow}
\usepackage{booktabs}
\usepackage{algorithmic}
\usepackage{algorithm}
\usepackage{xfrac}
\usepackage{fancyvrb}
\usepackage{balance}
\usepackage{adjustbox}

\usepackage[british]{babel}
\usepackage[none]{hyphenat}
\usepackage{parskip}

\usepackage{palatino}     
\usepackage{inconsolata}  

\usepackage{microtype}
\DisableLigatures[f]{encoding = *, family = * }

\graphicspath{{./}{./figures/}}

\DeclareMathOperator*{\argmin}{argmin}

\def\Vec#1{{\boldsymbol{#1}}}
\def\Mat#1{{\boldsymbol{#1}}}

\renewcommand{\baselinestretch}{1.1}\small\normalsize

\titleformat*{\section}{\bf\normalsize\large}
\titleformat*{\subsection}{\bf\normalsize}

\titlespacing{\section}{0pt}{\parskip}{0ex}
\titlespacing{\subsection}{0pt}{\parskip}{0ex}
\titlespacing{\subsubsection}{0pt}{\parskip}{0ex}

\begin{document}

\pagestyle{empty}
\lhead{}
\chead{}
\rhead{}
\lfoot{}
\cfoot{}
\rfoot{}

\begin{center}
{\Large\bf\doctitle}
\end{center}
\vspace{1ex}
\begin{center}

\begin{minipage}{0.5\textwidth}
{\large Conrad Sanderson~{$^{\dagger\diamond\ast}$} and Ryan Curtin~{$^{\ddagger\ast}$}}\\

\begin{small}
{$^\dagger$}  Data61, CSIRO, Australia\\
{$^\ddagger$} Symantec Corporation, USA\\
{$^\diamond$} University of Queensland, Australia\\
{$^\ast$}     Arroyo Consortium
\end{small}
\end{minipage}
\end{center}

\section*{Abstract}

\renewcommand{\baselinestretch}{1.0}\small\normalsize
\begin{small}
Modelling of multivariate densities is a core component in many signal processing, pattern recognition and machine learning applications.
The modelling is often done via Gaussian mixture models (GMMs), which use computationally expensive and potentially unstable training algorithms.
We provide an overview of a fast and robust implementation of GMMs in the C++ language,
employing multi-threaded versions of the Expectation Maximisation (EM) and \mbox{{\it k}-means} training algorithms.
Multi-threading is achieved through reformulation of the EM and \mbox{{\it k}-means} algorithms into a MapReduce-like framework.
Furthermore, the implementation uses several techniques to improve numerical stability and modelling accuracy.
We demonstrate that the multi-threaded implementation achieves a speedup of an order of magnitude on a recent 16 core machine,
and that it can achieve higher modelling accuracy than a previously well-established publically accessible implementation.
The multi-threaded implementation is included as a user-friendly class in recent releases of the open source Armadillo C++ linear algebra library.
The library is provided under the permissive Apache~2.0 license, allowing unencumbered use in commercial products.
\end{small}
\renewcommand{\baselinestretch}{1.1}\small\normalsize

\section{Introduction}

Modelling multivariate data through a convex mixture of Gaussians, also known as a Gaussian mixture model (GMM),
has many uses in fields such as signal processing, econometrics, pattern recognition, machine learning and computer vision.
Examples of applications include
multi-stage feature extraction for action recognition~\cite{Carvajal_2016a},
modelling of intermediate features derived from deep convolutional neural networks~\cite{Ge_ICIP_2015,Ge_2016,LeCun_Nature_2015},
classification of human epithelial cell images~\cite{Wiliem_PR_2014},
implicit sparse coding for face recognition~\cite{Wong_2014},
speech-based identity verification~\cite{Reynolds_2000},
and probabilistic foreground estimation for surveillance systems~\cite{Reddy_2013}.
GMMs are also commonly used as the emission distribution for hidden Markov models~\cite{Bilmes98}.

In the GMM approach, a distribution of samples (vectors) is modelled as:

\vspace{-3ex}
\begin{equation}
  p(\Vec{x} | \lambda) = \sum\nolimits_{g=1}^{N_G} w_g ~ {{\mathcal{N}}}( \Vec{x} | \Vec{\mu}_g, \Mat{\Sigma}_g )
  \label{eqn:gmm_prob}
\end{equation}%

\vspace{-2ex}
where $\Vec{x}$ is a $D$-dimensional vector,
$w_g$ is the weight for component $g$ (with constraints $\sum\nolimits_{g=1}^{N_G} w_g = 1$, $w_g \geq 0$),
and
${{\mathcal{N}}}( \Vec{x} | \Vec{\mu}, \Mat{\Sigma})$ is a $D$-dimensional Gaussian density function with mean $\Vec{\mu}$ and covariance matrix $\Mat{\Sigma}$:

\vspace{-3ex}
\begin{equation}
  {{\mathcal{N}}}( \Vec{x} | \Vec{\mu}, \Mat{\Sigma} )  = 
  \frac{1}{ (2\pi)^{\frac{D}{2}} | \Mat{\Sigma}|^{\frac{1}{2}} }
  \exp \left[ -\frac{1}{2} (\Vec{x}-\Vec{\mu})^\top \Mat{\Sigma}^{-1} (\Vec{x}-\Vec{\mu}) \right]
  \label{eqn:gaussian}
\end{equation}%

\vspace{-2ex}
where $|\Mat{\Sigma}|$ and $\Mat{\Sigma}^{-1}$ denote the determinant and inverse of $\Mat{\Sigma}$, respectively,
while $\Vec{x}^\top$ denotes the transpose of $\Vec{x}$.
The full parameter set can be compactly stated as {\small $\lambda = \{ w_g, \Vec{\mu}_g, \Mat{\Sigma}_g \}_{g=1}^{N_G}$},
where $N_G$ is the number of Gaussians.

Given a training dataset and a value for $N_G$,
the estimation of $\lambda$ is typically done through a
tailored instance of Expectation Maximisation (EM) algorithm~\cite{Dempster77, McLachlan-2008, Moon96, Redner84}.
The {\it k}-means algorithm~\cite{Bishop_2006,Duda01,Linde80} is also typically used for providing the initial estimate of $\lambda$ for the EM algorithm.
Choosing the optimal $N_G$ is data dependent and beyond the scope of this work; see~\cite{Hamerly_2003,Pelleg_2000} for example methods.

Unfortunately, GMM parameter estimation via the EM algorithm is computationally intensive
and can suffer from numerical stability issues.
Given the ever growing sizes of datasets and the need for fast, robust and accurate modelling of such datasets,
we have provided an open source implementation of multi-threaded (parallelised) versions 
of the \mbox{{\it k}-means} and EM algorithms.
In addition, core functions are recast in order to considerably reduce the likelihood of numerical instability due to floating point underflows and overflows.
The implementation is provided as a user-friendly class in recent releases of the cross-platform Armadillo C++ linear algebra library~\cite{Armadillo_JOSS_2016,Armadillo_PASC_2017}.
The library is licensed under the permissive Apache~2.0 license~\cite{Laurent_2008},
thereby allowing unencumbered use in commercial products.

We continue the paper as follows.
In Section~\ref{sec:param_em} we provide an overview of parameter estimation via the EM algorithm,
its reformulation for multi-threaded execution,
and approaches for improving numerical stability.
In Section~\ref{sec:param_km} we provide a summary of the {\it k}-means algorithm
along with approaches for improving its convergence and modelling accuracy.
The implementation in C++ is overviewed in~Section~\ref{sec:implementation},
where we list and describe the user accessible functions.
In Section~\ref{sec:eval}
we provide a demonstration that the implementation can achieve a speedup of an order of magnitude on a recent 16 core machine,
as well as obtain higher modelling accuracy than a previously well-established publically accessible implementation.

\section{Expectation Maximisation and Multi-Threading}
\label{sec:param_em}

The overall likelihood for a set of samples, $X=\{\Vec{x}_i\}_{i=1}^{N_V}$,
is found using $p(X | \lambda) = \prod\nolimits_{i=1}^{N_V} p(\Vec{x}_i | \lambda)$.
A~parameter set $\lambda$ that suitably models the underlying distribution of $X$ can be estimated using a particular instance of the Expectation Maximisation (EM) algorithm~\cite{Dempster77, McLachlan-2008, Moon96, Redner84}.
As its name suggests, the EM algorithm is comprised of iterating two steps: the {\it expectation} step, followed by the {\it maximisation} step.
GMM parameters generated by the previous iteration~($\lambda^{\textrm{old}}$) are used
by the current iteration to generate a new set of parameters~($\lambda^{\textrm{new}}$),
such that $p(X|\lambda^{\textrm{new}}) \geq p(X|\lambda^{\textrm{old}})$.

In a direct implementation of the EM algorithm specific to GMMs,
the estimated versions of the parameters ($\widehat{w}_g$, $\widehat{\Vec{\mu}}_g$, $\widehat{\Mat{\Sigma}}_g$)
within one iteration are calculated as follows:
\begin{eqnarray}
  l_{g,i}                  & = & \frac{w_g ~ {{\mathcal{N}}}( \Vec{x}_i | \Vec{\mu}_g, \Mat{\Sigma}_g )}{\sum\nolimits_{k=1}^{N_G} w_k ~ {{\mathcal{N}}}( \Vec{x}_i | \Vec{\mu}_k, \Mat{\Sigma}_k )}, \label{eqn:aposteriori} \\
  L_g                      & = & \sum\nolimits_{i=1}^{N_V} l_{g,i}, \label{eqn:em_sum_lhood} \\
  \widehat{w}_g            & = & \frac{L_g}{N_V},  \label{eqn:em_weight} \\
  \widehat{\Vec{\mu}}_g    & = & \frac{1}{L_g} \sum\nolimits_{i=1}^{N_V} \Vec{x}_i ~ l_{g,i}  \label{eqn:em_mean}, \\
  \widehat{\Mat{\Sigma}}_g & = & \frac{1}{L_g} \sum\nolimits_{i=1}^{N_V} (\Vec{x}_i - \widehat{\Vec{\mu}}_g)(\Vec{x}_i - \widehat{\Vec{\mu}}_g)^\top l_{g,i}. \label{eqn:em_cov}
\end{eqnarray}

Once the estimated parameters for all Gaussians are found, the parameters are updated,
{\small $\left\{ w_g, \Vec{\mu}_g, \Mat{\Sigma}_g \right\}_{g=1}^{N_G} = \left\{ \widehat{w}_g, \widehat{\Vec{\mu}}_g, \widehat{\Mat{\Sigma}}_g \right\}_{g=1}^{N_G}$},
and the iteration starts anew.
The process is typically repeated until the number of iterations has reached a pre-defined number,
and/or the increase in the overall likelihood after each iteration falls below a pre-defined threshold.

In Eqn.~(\ref{eqn:aposteriori}), $l_{g,i} \in [0,1]$ is the {a-posteriori} probability of Gaussian $g$ given $\Vec{x}_i$ and current parameters.
Thus the estimates $\widehat{\Vec{\mu}}_g$ and $\widehat{\Mat{\Sigma}}_g$ are weighted versions of the
sample mean and sample covariance, respectively.

Overall, the algorithm is a hill climbing procedure for maximising $p(X | \lambda)$.
While there are no guarantees that it will reach a global maximum, it is guaranteed to monotonically converge to a saddle point or a local maximum~\cite{Dempster77,Duda01,Mitchell97}.
The above implementation can also be interpreted as an unsupervised probabilistic clustering procedure,
with $N_G$ being the assumed number of clusters.
For a full derivation of the EM algorithm tailored to GMMs, the reader is directed to~\cite{Bilmes98,Redner84} or Appendix~A. 

\subsection{Reformulation for Multi-Threaded Execution}
\label{sec:param_em_parallel}

The EM algorithm is quite computationally intensive.
This is in large part due to the use of the {$\exp(\cdot)$} function, which needs to be applied numerous times for each and every sample.
Fortunately, CPUs with a multitude of cores are now quite common and accessible, allowing for multi-threaded (parallel) execution.

One approach for parallelisation is the MapReduce framework~\cite{MapReduce_2004},
where data is split into chunks and farmed out to separate workers for processing (mapping).
The results are then collected and combined (reduced) to produce the final result.
Below we provide a reformulation of the EM algorithm into a MapReduce-like framework.

As Eqn.~(\ref{eqn:aposteriori}) can be executed independently for each sample,
the summations in Eqns.~(\ref{eqn:em_sum_lhood}) and (\ref{eqn:em_mean}) can be split into separate sets of summations,
where the summation in each set can be executed independently and in parallel with other sets.
To allow similar splitting of the summation for calculating covariance matrices,
Eqn.~(\ref{eqn:em_cov}) needs to be rewritten into the following form:

\vspace{-3ex}
\begin{equation}
  \widehat{\Mat{\Sigma}}_g = \frac{1}{L_g} \left[ \sum\nolimits_{i=1}^{N_V} \Vec{x}_i \Vec{x}_i^\top l_{g,i} \right] - \widehat{\Vec{\mu}}_g \widehat{\Vec{\mu}}_g^\top.
\end{equation}

\vspace{-1ex}
The multi-threaded estimation of the parameters can now be formally stated as follows.
Given $N_T$ threads, the training samples are split into $N_T$ chunks, with each chunk containing approximately the same amount of samples.
For thread with index $t \in [1,N_T]$, the start index of the samples is denoted by $i^{[t]}_{\textrm{start}}$,
while the end index is denoted by $i^{[t]}_{\textrm{end}}$.
For~each thread $t$ and Gaussian $g \in [1,N_G]$, accumulators $\widetilde{L}_g^{[t]}$, $\widetilde{\Vec{\mu}}_g^{[t]}$ and $\widetilde{\Mat{\Sigma}}_g^{[t]}$
are calculated as follows:

\vspace{-5ex}
\begin{eqnarray}
  \widetilde{L}_g^{[t]}            & = & \sum\nolimits_{j = i^{[t]}_{\textrm{start}}}^{i^{[t]}_{\textrm{end}}} l_{g,j},                             \\ 
  \widetilde{\Vec{\mu}}_g^{[t]}    & = & \sum\nolimits_{j = i^{[t]}_{\textrm{start}}}^{i^{[t]}_{\textrm{end}}} l_{g,j} ~ \Vec{x}_j,                 \\ 
  \widetilde{\Mat{\Sigma}}_g^{[t]} & = & \sum\nolimits_{j = i^{[t]}_{\textrm{start}}}^{i^{[t]}_{\textrm{end}}} l_{g,j} ~ \Vec{x}_j \Vec{x}_j^\top.     
\end{eqnarray}%

\vspace{-2ex}
where $l_{g,j}$ is defined in Eqn.~(\ref{eqn:aposteriori}).

Once the accumulators for all threads are calculated,
for each Gaussian $g$ the reduction operation combines them to form the estimates of $\widehat{\Vec{\mu}}_g$ and $\widehat{\Mat{\Sigma}}_g$ as follows:

\vspace{-3ex}
\begin{eqnarray}
  L_g                      & = & \sum\nolimits_{t=1}^{N_T} \widetilde{L}_g^{[t]},                           \label{eqn:combined_L_g} \\
  \widehat{\Vec{\mu}}_g    & = & \frac{1}{L_g} \sum\nolimits_{t=1}^{N_T} \widetilde{\Vec{\mu}}_g^{[t]},                              \\
  \widehat{\Mat{\Sigma}}_g & = & \frac{1}{L_g} \sum\nolimits_{t=1}^{N_T} \widetilde{\Mat{\Sigma}}_g^{[t]} - \widehat{\Vec{\mu}}_g \widehat{\Vec{\mu}}_g^\top.
\end{eqnarray}

\vspace{-2ex}
The estimation of $\widehat{w}_g$ is as per Eqn.~(\ref{eqn:em_weight}), but using $L_g$ from Eqn.~(\ref{eqn:combined_L_g}).

\subsection{Improving Numerical Stability}

Due to the necessarily limited precision of numerical floating point representations~\cite{Goldberg_1991,Monniaux_2008},
the direct computation of Eqns.~(\ref{eqn:gmm_prob}) and (\ref{eqn:gaussian}) can quickly lead to numerical underflows or overflows,
which in turn lead to either poor models or a complete failure to estimate the parameters.
To address this problem, the following reformulation can be used.
First, logarithm version of Eqn.~(\ref{eqn:gaussian}) is taken:

\vspace{-3ex}
\begin{equation}
  \log {{\mathcal{N}}}( \Vec{x} | \Vec{\mu}, \Mat{\Sigma} )
  = -\left\{\frac{D}{2} \log \left( 2\pi \right) + \frac{1}{2} ~ \log ( |\Mat{\Sigma}| ) \right\}
    -\frac{1}{2} (\Vec{x}-\Vec{\mu})^\top \Mat{\Sigma}^{-1} (\Vec{x}-\Vec{\mu}),
\end{equation}

\vspace{-1ex}
which leads to the corresponding logarithm version of Eqn.~(\ref{eqn:gmm_prob}):

\vspace{-3ex}
\begin{equation}
\log \sum\nolimits_{g=1}^{N_G} w_g ~ {{\mathcal{N}}}( \Vec{x} ~|~ \Vec{\mu}_g, \Mat{\Sigma}_g )
=
\log \sum\nolimits_{g=1}^{N_G} \exp\left[ \log \left\{ w_g {{\mathcal{N}}}( \Vec{x} ~|~ \Vec{\mu}_g, \Mat{\Sigma}_g ) \right\} \right].
\label{eqn:log_gmm}
\end{equation}

\vspace{-2ex}
The right hand side of Eqn.~(\ref{eqn:log_gmm}) can be expressed as a repeated addition in the form of:

\vspace{-3ex}
\begin{equation}
\log\left( \exp\left[\log(a)\right] + \exp\left[\log(b)\right] \right),
\end{equation}

\vspace{-1ex}
which in turn can be rewritten in the form of:

\vspace{-3ex}
\begin{equation}
\log(a) + \log\left( 1 + \exp\left[ \log(b) - \log(a) \right] \right).
\end{equation}

\vspace{-1ex}
In the latter form, if we ensure that {$\log(a) \geq \log(b)$} (through swapping $\log(a)$ and $\log(b)$ when required),
the exponential will always produce values $\leq 1$ which helps to reduce the occurrence of overflows.
Overall, by keeping most of the computation in the {\it log} domain, 
both underflows and overflows are considerably reduced.

A further practical issue is the occurrence of \mbox{degenerate} or ill-conditioned covariance matrices,
stemming from either not enough samples with $l_{g,i} > 0$ contributing to the calculation of $\widehat{\Mat{\Sigma}}_g$ in Eqn.~(\ref{eqn:em_cov}),
or from too many samples which are essentially the same (ie.,~very low variance).
When the diagonal entries in a covariance matrix are too close to zero,
inversion of the matrix is unstable and can cause the calculated log-likelihood to become unreliable
or non-finite.
A straightforward and effective approach to address this problem is to place an artificial floor
on the diagonal entries in each covariance matrix after each EM iteration.
While the optimum value of the floor is data dependent, a small positive constant is typically sufficient to promote numerical stability
and convergence.

\section{Initialisation via Multi-Threaded {\it k}-Means}
\label{sec:param_km}

As a starting point, the initial means can be set to randomly selected training vectors,
the initial covariance matrices can be set equal to identity matrices, 
and the initial weights can be uniform.
However, the $\exp(\cdot)$ function as well as the matrix inverse in Eqn.~(\ref{eqn:gaussian}) are typically quite time consuming to compute.
In order to speed up training, the initial estimate of $\lambda$ is typically provided via the {\it k}-means clustering algorithm~\cite{Bishop_2006,Duda01,Kulis_2012}
which avoids such time consuming operations.

The baseline {\it k}-means clustering algorithm is a straightforward iterative procedure comprised of two steps:
(i)~calculating the distance from each sample to each mean,
and
(ii)~calculating the new version of each mean as the average of samples which were found to be the closest to the previous version of the corresponding mean.
The required number of iterations is data dependent,
but about 10 iterations are often sufficient to generate a good initial estimate of $\lambda$.

The $k$-means algorithm can be interpreted as a simplified version (or special case) of the EM algorithm for GMMs~\cite{Kulis_2012}.
Instead of each sample being assigned a set probabilities representing cluster membership (soft assignment),
each sample is assigned to only one cluster (hard assignment).
Furthermore, it can be assumed that the covariance matrix of each Gaussian is non-informative, diagonal, and/or shared across all Gaussians.
More formally, the estimation of model parameters is as per Eqns.~(\ref{eqn:em_weight}), (\ref{eqn:em_mean}) and (\ref{eqn:em_cov}), 
but $l_{g,i}$ is redefined~as:%
\begin{equation}
  l_{g,i} = \left\{
  \begin{array}{ll}
  1, & \mbox{if} ~ g = \argmin\limits_{k=1, \cdots, N_G} \operatorname{dist}(\Vec{\mu}_k, \Vec{x}_i) \\
  0, & \mbox{otherwise}.
  \end{array}
  \right.
  \label{eqn:binary_likelihood}
\end{equation}
where {$\operatorname{dist}(\Vec{a}, \Vec{b})$} is a distance metric.
Apart from this difference, the parameter estimation is the same as for EM.
As such, multi-threading is achieved as per Section~\ref{sec:param_em_parallel}.

We note that it is possible to implement the \mbox{{\it k}-means} algorithm is a multitude of ways,
such as the cluster splitting LBG algorithm~\cite{Linde80},
or use an elaborate strategy for selecting the initial means~\cite{Arthur_2007}.
While there are also alternative and more complex implementations offering relatively fast execution~\cite{Elkan_2003},
we have elected to adapt the baseline \mbox{{\it k}-means} algorithm due to its straightforward amenability to multi-threading.

\subsection{Issues with Modelling Accuracy and Convergence}

A typical and naive choice for the distance in~Eqn.~(\ref{eqn:binary_likelihood})
is the squared Euclidean distance, \mbox{$\operatorname{dist}(\Vec{a}, \Vec{b}) = \| \Vec{a} - \Vec{b} \|^{2}_{2}$}.
However, for multivariate datasets formed by combining data from various sensors, there is a notable downside to using the Euclidean distance.
When one of the dimensions within the data has a much larger range than the other dimensions,
it will dominate the contribution to the overall distance, with the other dimensions effectively ignored.
This can adversely skew the initial parameter estimates, easily leading to poor initial conditions for the EM algorithm.
This in turn can lead to poor modelling, as the EM algorithm is only guaranteed to reach a local maximum~\cite{Dempster77,Duda01,Mitchell97}.
To address this problem, the squared Mahalanobis distance can be used~\cite{Bishop_2006,Duda01}:
\begin{equation}
\operatorname{dist}(\Vec{a}, \Vec{b}) = (\Vec{a} - \Vec{b})^\top \Mat{\Sigma}^{-1}_{\mathrm{global}} (\Vec{a} - \Vec{b}),
\end{equation}
\noindent where $\Mat{\Sigma}_{\mathrm{global}}$ is a global covariance matrix, estimated from all available training data.
To maintain efficiency, $\Mat{\Sigma}_{\mathrm{global}}$ is typically diagonal,
which makes calculating its inverse straightforward (ie., reciprocals of the values on the main diagonal).

In practice it is possible that while iterating at least one of the means has no vectors assigned to it,
becoming a ``dead'' mean.
This might stem from an unfortunate starting point, 
or specifying a relatively large value for $N_G$ for modelling a relatively small dataset.
As such, an additional heuristic is required to attempt to resolve this situation.
An effective approach for resurrecting a ``dead'' mean is to make it equal to one of the vectors
that has been assigned to the most ``popular'' mean,
where the most ``popular'' mean is the mean that currently has the most vectors assigned to it.

\newpage
\section{Implementation in C++}
\label{sec:implementation}

We have provided a numerical implementation of Gaussian Mixture Models in the C++ language
as part of recent releases of the open source Armadillo C++ linear algebra library~\cite{Armadillo_JOSS_2016}. 
The library is available under the permissive Apache~2.0 license~\cite{Laurent_2008},
and can be obtained from
\href{http://arma.sourceforge.net}{http://arma.sourceforge.net}.
To considerably reduce execution time,
the implementation contains multi-threaded versions of the EM and {\it k}-means training algorithms
(as overviewed in Sections~\ref{sec:param_em} and~\ref{sec:param_km}).
Implementation of multi-threading is achieved with the aid of OpenMP {\it pragma} directives~\cite{OpenMP_2007}.

There are two main choices for the type of covariance matrix $\Mat{\Sigma}$: full and diagonal.
While full covariance matrices have more capacity for modelling data,
diagonal covariance matrices provide several practical advantages:
\begin{enumerate}[{\bf (i)}]
\item
the computationally expensive (and potentially unstable) matrix inverse operation in Eqn.~(\ref{eqn:gaussian})
is reduced to simply to taking the reciprocals of the diagonal elements,

\item
the determinant operation is considerably simplified to taking the product of the diagonal elements,

\item
diagonal covariance matrices contain fewer parameters that need to be estimated, and hence require fewer training samples~\cite{Duda01}.
\end{enumerate}

Given the above practical considerations, the implementation uses diagonal covariance matrices.
We note that diagonal covariance GMMs with $N_G > 1$ can model distributions of samples with correlated elements,
which in turn suggests that full covariance GMMs can be approximated using diagonal covariance GMMs with a larger number of Gaussians~\cite{Reynolds_2000}.

\subsection{User Accessible Classes and Functions}

The implementation is provided as two user-friendly classes within the {\it arma} namespace:
{\it\bfseries gmm\_diag} and {\it\bfseries fgmm\_diag}.
The~former uses double precision floating point values, while the latter uses single precision floating point values.
For an instance of the double precision {\it gmm\_diag} class named as~{\bf M},
its member functions and variables are listed below.
The interface allows the user full control over the parameters for GMM fitting,
as well as easy and flexible access to the trained model.
Figure~\ref{fig:example_usage} contains a complete C++ program which demonstrates usage of the {\it gmm\_diag} class.

In the description below, all vectors and matrices refer to corresponding objects from the Armadillo library;
scalars have the type {\it double},
matrices have the type {\it mat},
column vectors have the type {\it vec},
row vectors have the type {\it rowvec},
row vectors of unsigned integers have the type {\it urowvec},
and indices have the type {\it uword} (representing an unsigned integer).
When using the single precision {\it fgmm\_diag} class,
all vector and matrix types have the {\it f} prefix (for example, {\it fmat}),
while scalars have the type {\it float}.
The word ``heft'' is explicitly used in the classes as a shorter version of ``weight'', while keeping the same meaning with the context of GMMs.

\begin{small}
\begin{enumerate}[{$\bullet$}]
\itemsep 1ex

\item
{\bf M.log\_p(V)}\\
return a scalar (double precision floating point value) representing the log-likelihood of column vector {\bf V}

\item
{\bf M.log\_p(V, g)}\\
return a scalar (double precision floating point value) representing the log-likelihood of column vector {\bf V},
according to Gaussian with index~{\bf g} (specified as an unsigned integer of type {\it uword})

\item
{\bf M.log\_p(X)}\\
return a row vector (of type {\it rowvec}) containing log-likelihoods of each column vector in matrix {\bf X}

\item
{\bf M.log\_p(X, g)}\\
return a row vector (of type {\it rowvec}) containing log-likelihoods of each column vector in matrix {\bf X},
according to Gaussian with index {\bf g}  (specified as an unsigned integer of type {\it uword})

\item
{\bf M.avg\_log\_p(X)}\\
return a scalar (double precision floating point value) representing the average log-likelihood of all column vectors in matrix {\bf X}

\item
{\bf M.avg\_log\_p(X, g)}\\
return a scalar (double precision floating point value) representing the average log-likelihood of all column vectors in matrix {\bf X},
according to Gaussian with index~{\bf g}  (specified as an unsigned integer of type {\it uword})

\item
{\bf M.assign(V, dist\_mode)}\\
return an unsigned integer (of type {\it uword}) representing the index of the
closest mean (or Gaussian) to vector {\bf V}; the parameter {\bf dist\_mode} is one
of:
\begin{small}
\begin{enumerate}[{\bf {~eucl\_dist}}]
\item
Euclidean distance (takes only means into account)
\end{enumerate}

\begin{enumerate}[{\bf {prob\_dist}}]
\item
probabilistic ``distance'', defined as the inverse likelihood (takes into account means, covariances and hefts)
\end{enumerate}
\end{small}


\item
{\bf M.assign(X, dist\_mode)}\\
return a row vector of unsigned integers (of type {\it urowvec}) containing the indices of the closest means (or Gaussians) to each column vector in matrix {\bf X};
parameter {\bf dist\_mode} is {\bf eucl\_dist} or {\bf prob\_dist}, as per the {\bf .assign()} function above

\item
{\bf M.raw\_hist(X, dist\_mode)}\\
return a row vector of unsigned integers (of type {\it urowvec}) representing the raw histogram of counts;
each entry is the number of counts corresponding to a Gaussian;
each count is the number times the corresponding Gaussian was the closest to each column vector in matrix {\bf X};
parameter {\bf dist\_mode} is {\bf eucl\_dist} or {\bf prob\_dist}, as per the {\bf .assign()} function above

\item
{\bf M.norm\_hist(X, dist\_mode)}\\
similar to the {\bf .raw\_hist()} function above; return a row vector (of type {\it rowvec}) containing normalised counts; the vector sums to one;
parameter {\bf dist\_mode} is either {\bf eucl\_dist} or {\bf prob\_dist}, as per the {\bf .assign()} function above

\item
{\bf M.generate()}\\
return a column vector (of type {\it vec}) representing a random sample generated according to the model's parameters

\item
{\bf M.generate(N)}\\
return a matrix (of type {\it mat}) containing {\bf N} column vectors, with each vector representing a random sample generated according to the model's parameters

\item
{\bf M.n\_gaus()}\\
return an unsigned integer (of type {\it uword}) containing the number of means/Gaussians in the model

\item
{\bf M.n\_dims()}\\
return an unsigned integer (of type {\it uword}) containing the dimensionality of the means/Gaussians in the model

\item
{\bf M.reset(n\_dims, n\_gaus)}\\
set the model to have dimensionality {\bf n\_dims}, with {\bf n\_gaus} number of Gaussians, specified as unsigned integers of type {\it uword};
all the means are set to zero, all diagonal covariances are set to one, and all the hefts (weights) are set to be uniform

\item
{\bf M.save(filename)}\\
save the model to a file and return a {\it bool} indicating either success ({\it true}) or failure ({\it false})

\item
{\bf M.load(filename)}\\
load the model from a file and return a {\it bool} indicating either success ({\it true}) or failure ({\it false})

\item
{\bf M.means}\\
read-only matrix (of type {\it mat}) containing the means (centroids), stored as column vectors

\item
{\bf M.dcovs}\\
read-only matrix (of type {\it mat}) containing the diagonal covariances, with the set of diagonal covariances for each Gaussian stored as a column vector

\item
{\bf M.hefts}\\
read-only row vector (of type {\it rowvec}) containing the hefts (weights)

\item
{\bf M.set\_means(X)}\\
set the means (centroids) to be as specified in matrix {\bf X} (of type {\it mat}), with each mean (centroid) stored as a column vector;
the number of means and their dimensionality must match the existing model

\item
{\bf M.set\_dcovs(X)}\\
set the diagonal covariances to be as specified in matrix~{\bf X} (of type {\it mat}), with the set of diagonal covariances for each Gaussian stored as a column vector;
the number of diagonal covariance vectors and their dimensionality must match the existing model

\item
{\bf M.set\_hefts(V)}\\
set the hefts (weights) of the model to be as specified in row vector {\bf V} (of type {\it rowvec});
the number of hefts must match the existing model

\pagebreak
\item
{\bf M.set\_params(means, dcovs, hefts)}\\
set all the parameters at the same time, using matrices denoted as {\bf means} and {\bf dcovs} as well as the row vector denoted as {\bf hefts};
the layout of the matrices and vectors is as per the {\bf .set\_means()}, {\bf .set\_dcovs()} and {\bf .set\_hefts()} functions above;
the number of Gaussians and dimensionality can be different from the existing model

\item
{\bf M.learn(data, n\_gaus, dist\_mode, seed\_mode, km\_iter, em\_iter, var\_floor, print\_mode)}\\
learn the model parameters via the {\it k}-means and/or EM algorithms,
and return a boolean value, with {\it true} indicating success, and {\it false} indicating failure;
the parameters have the following meanings:

\begin{small}
\begin{enumerate}[{-}]
\itemsep 0.5ex
\item
{\bf data}\\
matrix (of type {\it mat}) containing training samples; each sample is stored as a column vector

\item
{\bf n\_gaus}\\
set the number of Gaussians to {\bf n\_gaus};
to help convergence, it is recommended that the given {\bf data} matrix (above)
contains at least 10 samples for each Gaussian

\item
{\bf dist\_mode}\\
specifies the distance used during the seeding of initial means and \mbox{{\it k}-means} clustering:
\begin{small}
\begin{enumerate}[{\bf eucl\_dist}]
\item Euclidean distance
\end{enumerate}

\begin{enumerate}[{\bf ~{maha\_dist}}]
\item Mahalanobis distance, which uses a global diagonal covariance matrix estimated from the given training samples
\end{enumerate}
\end{small}


\item
{\bf seed\_mode}\\
specifies how the initial means are seeded prior to running \mbox{{\it k}-means} and/or EM algorithms:


\vspace*{0.5em}
\begin{tabular}{rl}
{\bf keep\_existing} & keep the existing model (do not modify the means, covariances and hefts) \\
{\bf static\_subset} & a subset of the training samples (repeatable) \\
{\bf random\_subset} & a subset of the training samples (random) \\
{\bf static\_spread} & a maximally spread subset of training samples (repeatable) \\
{\bf random\_spread} & a maximally spread subset of training samples (random start) \\
\end{tabular}
\vspace*{0.5em}

Note that seeding the initial means with {\bf static\_spread} and {\bf random\_spread}
can be more time consuming than with {\bf static\_subset} and {\bf random\_subset};
these seed modes are inspired by the so-called {\it k-means++} approach~\cite{Arthur_2007}, with the aim to improve clustering quality.
\vspace{1ex}

\item
{\bf km\_iter}\\
the maximum number of iterations of the {\it k}-means algorithm; this is data dependent, but typically 10 iterations are sufficient

\item
{\bf em\_iter}\\
the maximum number of iterations of the EM algorithm; this is data dependent, but typically 5 to 10 iterations are sufficient

\item
{\bf var\_floor}\\
the variance floor (smallest allowed value) for the diagonal covariances; setting this to a small non-zero value can help with convergence and/or better quality parameter estimates

\item
{\bf print\_mode}\\
boolean value (either {\it true} or {\it false}) which enables/disables the printing of progress during the {\it k}-means and EM algorithms 

\end{enumerate}
\end{small}

\end{enumerate}
\end{small}

\begin{figure}[!h]
\vspace{1ex}
\hrule
\vspace{1ex}
\centering
\begin{adjustbox}{minipage=\columnwidth,scale={1}{1.1}}
\begin{Verbatim}[fontsize=\footnotesize]
#include <armadillo>

using namespace arma;

int main()
  {
  // create synthetic data containing
  // 2 clusters with normal distribution
  
  uword d = 5;       // dimensionality
  uword N = 10000;   // number of samples (vectors)
  
  mat data(d, N, fill::zeros);
  
  vec mean1 = linspace<vec>(1,d,d);
  vec mean2 = mean1 + 2;
  
  uword i = 0;
  
  while(i < N)
    {
    if(i < N)  { data.col(i) = mean1 + randn<vec>(d); ++i; }
    if(i < N)  { data.col(i) = mean1 + randn<vec>(d); ++i; }
    if(i < N)  { data.col(i) = mean2 + randn<vec>(d); ++i; }
    }
  
  // model the data as a diagonal GMM with 2 Gaussians
  
  gmm_diag model;
  
  bool status = model.learn(data, 2, maha_dist, random_subset,
                            10, 5, 1e-10, true);
  
  if(status == false)  { cout << "learning failed" << endl; }
  
  model.means.print("means:");
  
  double overall_likelihood = model.avg_log_p(data);
  
  rowvec     set_likelihood = model.log_p( data.cols(0,9) );
  double  scalar_likelihood = model.log_p( data.col(0)    );
  
  uword   gaus_id  = model.assign( data.col(0),    eucl_dist );
  urowvec gaus_ids = model.assign( data.cols(0,9), prob_dist );
  
  urowvec histogram1 = model.raw_hist (data, prob_dist);
   rowvec histogram2 = model.norm_hist(data, eucl_dist);
  
  model.save("my_model.gmm");
  
  mat modified_dcovs = 2 * model.dcovs;
  
  model.set_dcovs(modified_dcovs);
  
  return 0;
  }

\end{Verbatim}
\end{adjustbox}
\hrule
\vspace{0.5ex}
\caption
  {
  An example C++ program which demonstrates usage of a subset of functions available in the {\it\bfseries gmm\_diag} class.
  }
\label{fig:example_usage}
\end{figure}

\clearpage
\section{Evaluation}
\label{sec:eval}

\subsection{Speedup from Multi-Threading}

To demonstrate the achievable speedup with the multi-threaded versions of the EM and {\it k}-means algorithms,
we trained a GMM with 100 Gaussians on a recent 16 core machine using a synthetic dataset comprising 1,000,000 samples with 100 dimensions.
10 iterations of the {\it k}-means algorithm and 10 iterations of the EM algorithm were used.
The samples were stored in double precision floating point format, resulting in a total data size of approximately 762~Mb.

\begin{figure*}[!b]
\centering
\begin{minipage}{\textwidth}
  \centering
  \begin{minipage}{0.5\textwidth}
    \centering
    \includegraphics[width=1.1\textwidth]{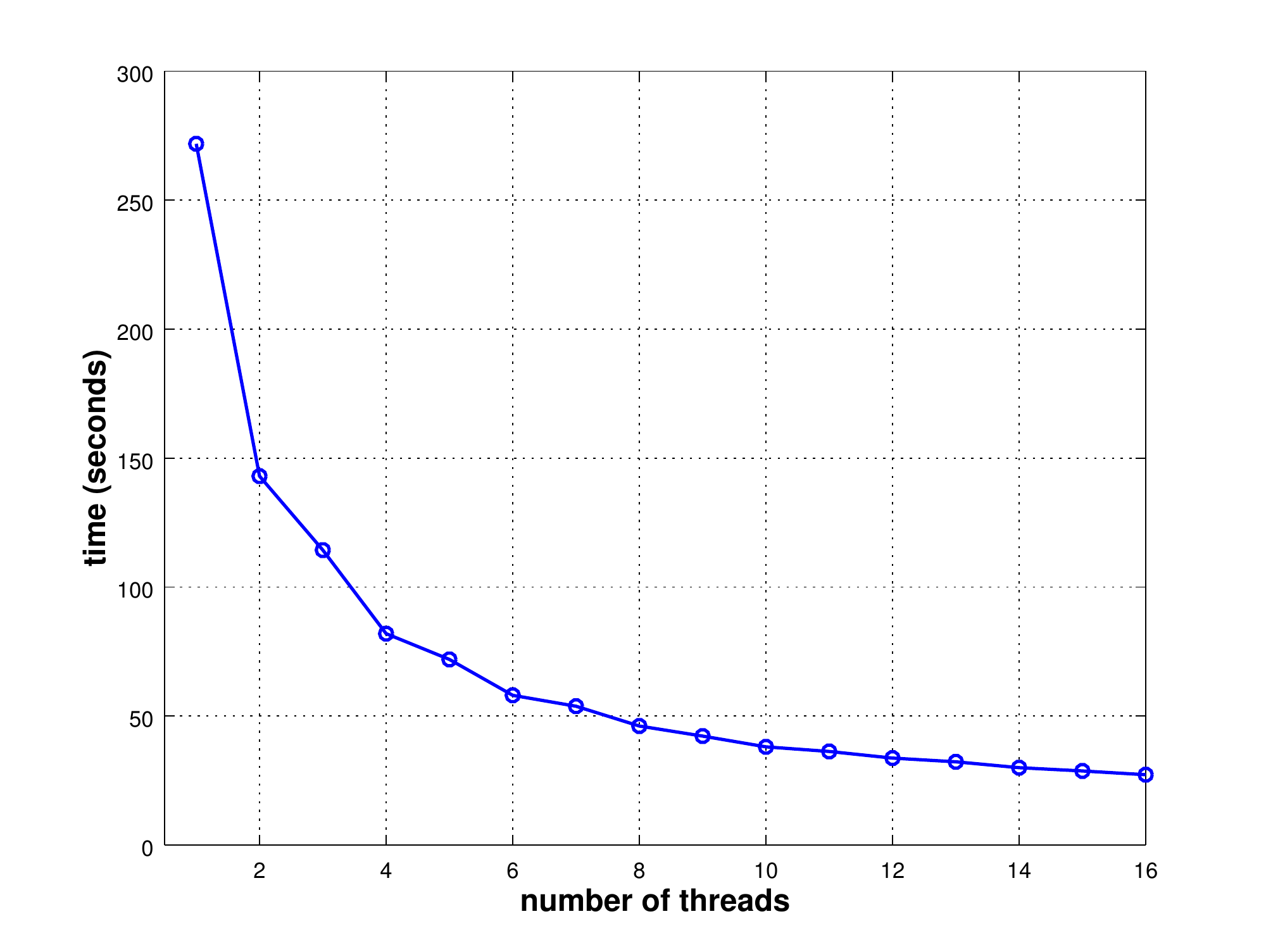}\\
    {(a)}
  \end{minipage}%
  \begin{minipage}{0.5\textwidth}
    \centering
    \includegraphics[width=1.1\textwidth]{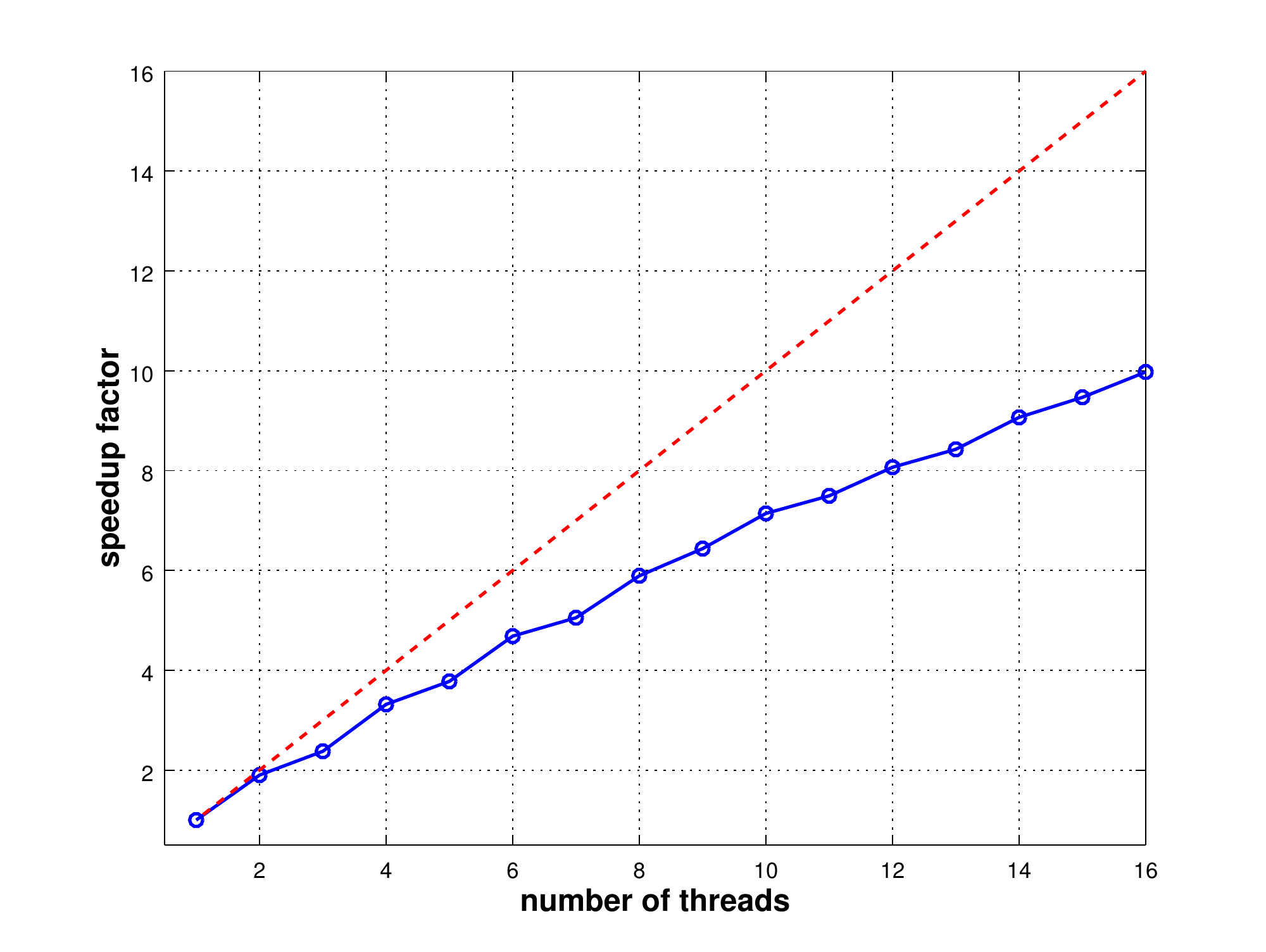}\\
    {(b)}
  \end{minipage}
\end{minipage}
\caption
  {
  Execution characteristics for training a 100 component GMM
  to model a synthetic dataset comprising 1,000,000 samples with 100 dimensions,
  using 10 iterations of the {\it k}-means algorithm and 10 iterations of the EM algorithm:
  {\bf (a)}~total~time~taken depending on the number of threads;
  {\bf (b)}~corresponding speedup factor compared to using one thread (blue line), and idealised linear speedup under the assumption of no overheads and no memory access contention (red dotted line).
  The modelling was done on a machine with dual Intel Xeon E5-2620-v4 CPUs, providing 16 independent processing cores running at 2.1~GHz.
  Compilation was done with the GCC 5.4 C++ compiler with the following configuration options: \texttt{-O3 -march=native -fopenmp}.
  }
\label{fig:speedup}
\end{figure*}

Figure~\ref{fig:speedup} shows that a speedup of an order of magnitude is achieved when all 16 cores are used.
Specifically, for the synthetic dataset used in this demonstration,
the training time was reduced from approximately 272 seconds to about 27 seconds.
In each case, the {\it k}-means algorithm took approximately 30\% of the total training time.

We note that the overall speedup is below the idealised linear speedup.
This is likely due to overheads related to OpenMP and reduction operations described in Section~\ref{sec:param_em_parallel},
as well as memory access contention, stemming from concurrent access to memory by multiple cores~\cite{McCool_2012}.

\subsection{Comparison with Full-Covariance GMMs in MLPACK}

In order to validate our intuition that a diagonal GMM is a good choice instead
of the significantly more complex problem of estimating GMMs with full covariance matrices,
we compare the {\it gmm\_diag} class (described in Section~\ref{sec:implementation})
against the full-covariance GMM implementation in the well-established MLPACK C++ machine learning library~\cite{Curtin_2013}.

We selected common datasets from the UCI machine learning dataset repository~\cite{Lichman_2013},
and trained both diagonal and full-covariance GMMs on these datasets.
The number of Gaussians was chosen according to the original source of each dataset;
where possible, 3 times the number of classes in the dataset was used.
In some cases, small amounts of Gaussian noise was added to the dataset to ensure
training stability of the full-covariance GMMs.
Both implementations used 10 iterations of {\it k}-means for initialisation,
followed by running the EM algorithm until convergence or reaching a maximum of 250 iterations.
The entire fitting procedure was repeated 10 times, each time with a different random starting point.

The results are given in Table \ref{tab:results}, which shows the best log-likelihood of the 10 runs,
the average wall-clock runtime for the fitting,
as well as dataset information
(number of samples, dimensionality, and number of Gaussians used for modelling).
We can see that the diagonal GMM implementation in 
the {\it gmm\_diag} class provides speedups from one to two orders-of-magnitude
over the full-covariance implementation in \mbox{MLPACK}.
Furthermore, in most cases there is no significant loss in goodness-of-fit (as measured by log-likelihood).
In several cases ({\it winequality}, {\it phy}, {\it covertype}, {\it pokerhand})
the log-likelihood is notably higher for the {\it gmm\_diag} class;
we conjecture that in these cases the diagonal covariance matrices are acting as a form of regularisation to reduce overfitting~\cite{Bishop_2006}.

\begin{table*}[!tb]
\centering
\small
\begin{tabular}{|l|c|c|c|c|c|c|c|c|}
\hline
\multirow{2}{*}{\bfseries dataset} & {\bfseries num.}    & {\bfseries num.} & {\bfseries num.}  & {\bfseries MLPACK}   & {\bfseries {gmm\_diag}} & $\sfrac{\mbox{\bfseries MLPACK}}{\mbox{\bfseries gmm\_diag}}$ & {\bfseries MLPACK}             & {\bfseries {gmm\_diag}}        \\
                                   & {\bfseries samples} & {\bfseries dims} & {\bfseries Gaus.} & {\bfseries fit time} & {\bfseries fit time}    & {\bfseries fit time ratio}                                    & {$\mathbf \log~p(X|\lambda) $} & {$\mathbf \log~p(X|\lambda) $} \\
\hline
  cloud       & {\tt ~~~~2,048} & {\tt 10} & {\tt ~5} & {\tt ~~~~1.50s} &  {\tt\bfseries ~0.14s} & {\tt ~10.7} & {\tt~-59.98{\tiny$\times$}10$^{\mathtt 3}$} & {\tt          ~-64.12{\tiny$\times$}10$^{\mathtt 3}$} \\
  ozone       & {\tt ~~~~2,534} & {\tt 72} & {\tt ~6} & {\tt ~~~~8.59s} &  {\tt\bfseries ~0.10s} & {\tt ~85.9} & {\tt-226.13{\tiny$\times$}10$^{\mathtt 3}$} & {\tt          -307.95{\tiny$\times$}10$^{\mathtt 3}$} \\
  winequality & {\tt ~~~~6,497} & {\tt 11} & {\tt 30} & {\tt ~~~16.10s} &  {\tt\bfseries ~0.68s} & {\tt ~23.7} & {\tt~-47.12{\tiny$\times$}10$^{\mathtt 3}$} & {\tt\bfseries ~-15.85{\tiny$\times$}10$^{\mathtt 3}$} \\
  corel       & {\tt ~~~37,749} & {\tt 32} & {\tt 50} & {\tt ~~544.62s} &  {\tt\bfseries ~4.55s} & {\tt 119.7} & {\tt~~+4.52{\tiny$\times$}10$^{\mathtt 6}$} & {\tt          ~~+4.44{\tiny$\times$}10$^{\mathtt 6}$} \\
  birch3      & {\tt ~~100,000} & {\tt ~2} & {\tt ~6} & {\tt ~~~18.13s} &  {\tt\bfseries ~2.39s} & {\tt ~~7.6} & {\tt~~-2.70{\tiny$\times$}10$^{\mathtt 6}$} & {\tt          ~~-2.71{\tiny$\times$}10$^{\mathtt 6}$} \\
  phy         & {\tt ~~150,000} & {\tt 78} & {\tt 30} & {\tt ~3867.12s} &  {\tt\bfseries 29.25s} & {\tt 132.2} & {\tt~~-2.10{\tiny$\times$}10$^{\mathtt 7}$} & {\tt\bfseries ~~-1.88{\tiny$\times$}10$^{\mathtt 7}$} \\
  covertype   & {\tt ~~581,012} & {\tt 55} & {\tt 21} & {\tt 10360.53s} &  {\tt\bfseries 64.83s} & {\tt 159.8} & {\tt~~-9.46{\tiny$\times$}10$^{\mathtt 7}$} & {\tt\bfseries ~~-6.90{\tiny$\times$}10$^{\mathtt 7}$} \\
  pokerhand   & {\tt 1,000,000} & {\tt 10} & {\tt 25} & {\tt ~3653.94s} &  {\tt\bfseries 55.85s} & {\tt ~65.4} & {\tt~~-1.90{\tiny$\times$}10$^{\mathtt 7}$} & {\tt\bfseries ~~-1.68{\tiny$\times$}10$^{\mathtt 7}$} \\
\hline
\end{tabular}
\vspace{1ex}
\caption
  {
  Comparison of fitting time (seconds) and goodness-of-fit (as measured by log-likelihood) using full covariance GMMs from the MLPACK library~\cite{Curtin_2013}
  against diagonal GMMs in the {\it gmm\_diag} class,
  on common datasets from the UCI machine learning dataset repository~\cite{Lichman_2013}.
  The lower the fitting time, the better.
  The higher the $\log p(X|\lambda)$, the better.
  }
\label{tab:results}
\end{table*}

\section{Conclusion}

In this paper we have demonstrated a multi-threaded and robust implementation
of Gaussian Mixture Models in the C++ language.
Multi-threading is achieved through reformulation of the Expectation-Maximisation and {\it k}-means algorithms into a MapReduce-like framework.
The implementation also uses several techniques to improve numerical stability and improve modelling accuracy.
We demonstrated that the implementation achieves a speedup of an order of magnitude on a recent 16 core machine,
and that it can achieve higher modelling accuracy than a previously well-established publically accessible implementation.
The multi-threaded implementation is released as open source software
and included in recent releases of the cross-platform Armadillo C++ linear algebra library.
The library is provided under the permissive Apache~2.0 license, allowing unencumbered use in commercial products.

\appendix
\section*{Appendix A: Abridged Derivation of the EM Algorithm for Gaussian Mixture Models}
\label{app:em_algorithm}

In the Gaussian Mixture Model (GMM) approach, the distribution of samples (vectors) is modelled as:
\begin{equation}
	p(\Vec{x} | \Theta) = \sum\nolimits_{m=1}^{M} w_m p(\Vec{x}| \theta_m)
	\label{eqn:mixture_fn}
\end{equation}

\noindent
where $\Vec{x}$ is a $D$-dimensional vector,
$w_m$ is a weight (with constraints $\sum\nolimits_{m=1}^{M} w_m = 1$, $w_m \geq 0$),
and
$p(\Vec{x}| \theta_m)$ is a multivariate Gaussian density function with parameter set $\theta_m = \{ \Vec{\mu}_m, \Mat{\Sigma}_m \}$:
\begin{equation}
	p(\Vec{x}| \theta_m) =  {{\mathcal{N}}}( \Vec{x} | \Vec{\mu}_m, \Mat{\Sigma}_m )  = 
		\frac{1}{ (2\pi)^{\frac{D}{2}} | \Mat{\Sigma}_m|^{\frac{1}{2}} }
		\exp \left[ -\frac{1}{2} (\Vec{x}-\Vec{\mu}_m)^T \Mat{\Sigma}_m^{-1} (\Vec{x}-\Vec{\mu}_m) \right]
\end{equation}%

\noindent
where $\Vec{\mu}_m$ is the mean vector and $\Mat{\Sigma}_m$ is the covariance matrix.
Thus the complete parameter set for Eqn.~(\ref{eqn:mixture_fn}) is expressed as $\Theta = \{w_m, \theta_m\}_{m=1}^{M}$.
Given a set of training samples, $X=\{\Vec{x}_i\}_{i=1}^{N}$,
we need to find $\Theta$ that suitably models the underlying distribution.
Stated more formally, we need to find $\Theta$ that maximises the following likelihood function:
\begin{equation}
	p(X | \Theta) = \prod\nolimits_{i=1}^{N} p(\Vec{x}_i | \Theta)
	\label{eqn:lhood_fn}
\end{equation}

The Expectation-Maximisation (EM) algorithm~\cite{Dempster77, McLachlan-2008, Moon96, Redner84} is an iterative likelihood function optimisation technique,
often used in the pattern recognition and machine learning~\cite{Bishop_2006,Duda01}.
It is a general method for finding the maximum-likelihood estimate of the parameters of an assumed distribution,
when either the training data is incomplete or has missing values, or when the likelihood function can be made analytically tractable
by assuming the existence of (and values for) {\it missing} data.

To apply the EM algorithm to finding $\Theta$, we must first assume that our training data $X$ is incomplete
and assume the existence of missing data $Y = \{y_i\}_{i=1}^{N}$,
where each $y_i$ indicates the mixture component that ``generated'' the corresponding $\Vec{x}_i$.
Thus $y_i \in [1,M] ~ \forall ~ i$ and $y_i = m$ if the $i$-th feature vector ($\Vec{x}_i$) was ``generated'' by the $m$-th component.
If we know the values for $Y$, then Eqn.~(\ref{eqn:lhood_fn}) can be modified to:
\begin{equation}
	p(X,Y | \Theta) = \prod\nolimits_{i=1}^{N} w_{y_i} p(\Vec{x}_i| \theta_{y_i})
	\label{eqn:lhood_fn_modified}
\end{equation}

\noindent
As its name suggests, the EM algorithm is comprised of two steps which are iterated: (i)~expectation, followed by (ii)~maximisation.
In the expectation step, the expected value of the complete data log-likelihood, $\log p(X,Y | \Theta)$,
is found with respect to the unknown data $Y = \{y_i\}_{i=1}^{N}$ given training data $X=\{\Vec{x}_i\}_{i=1}^{N}$ and current parameter estimates,
$\Theta^{[k]}$ (where $k$ indicates the iteration number):
\begin{equation}
	Q(\Theta, \Theta^{[k]}) = E \left[ \log p(X,Y | \Theta) ~|~ X,\Theta^{[k]} \right]
	\label{eqn:q_fn}
\end{equation}

\noindent
Since $Y$ is a random variable with distribution $p(\mbox{\boldmath $y$}|X,\Theta^{[k]})$, Eqn.~(\ref{eqn:q_fn}) can be written as:
\begin{equation}
	Q(\Theta, \Theta^{[k]}) = \int_{\mbox{\boldmath $y$} \in \Upsilon}
								\log p(X,\mbox{\boldmath $y$} | \Theta) ~ p(\mbox{\boldmath $y$}|X,\Theta^{[k]}) ~~ d \mbox{\boldmath $y$}
	\label{eqn:q_fn2}
\end{equation}

\noindent
where \mbox{\boldmath $y$} is an instance of the missing data and $\Upsilon$ is the space of values \mbox{\boldmath $y$} can take on.
The maximisation step then maximises the expectation:
\begin{equation}
	\Theta^{[k+1]} = \arg \max_{\Theta} Q(\Theta, \Theta^{[k]})
	\label{eqn:maximize}
\end{equation}

\noindent
The expectation and maximisation steps are iterated until convergence,
or when the increase in likelihood falls below a pre-defined threshold.
As can be seen in Eqn.~(\ref{eqn:q_fn2}), we require $p(\mbox{\boldmath $y$}|X,\Theta^{[k]})$.
We can define it as follows:
\begin{equation}
	p(\mbox{\boldmath $y$}|X,\Theta^{[k]}) = \prod\nolimits_{i=1}^{N} p(y_i | \Vec{x}_i, \Theta^{[k]})
\end{equation}

\noindent
Given initial parameters%
\footnote{Parameters for $k=0$ can be found via the {\it k}-means algorithm~\cite{Bishop_2006,Duda01,Linde80} (see also Section \ref{sec:param_km}).}
~$\Theta^{[k]}$,
we can compute $p(\Vec{x}_i | \theta_m^{[k]})$.
Moreover, we can interpret the mixing weights ($w_m$) as {a-priori} probabilities of each mixture component, ie., $w_m = p(m | \Theta^{[k]})$.
Hence we can apply Bayes' rule~\cite{Duda01} to obtain:
\begin{eqnarray}
	p(y_i | \Vec{x}_i, \Theta^{[k]}) & = & \frac{ p(\Vec{x}_i | \theta_{y_i}^{[k]}) p(y_i | \Theta^{[k]}) }{ p(\Vec{x}_i | \Theta^{[k]}) } \\
	~ & = & \frac{ p(\Vec{x}_i | \theta_{y_i}^{[k]}) p(y_i | \Theta^{[k]}) }{ \sum\nolimits_{n=1}^{M} p(\Vec{x}_i | \theta_{n}^{[k]}) p(n | \Theta^{[k]})}
	\label{eqn:p_yi}
\end{eqnarray}

\noindent
Expanding Eqn.~(\ref{eqn:q_fn2}) yields:
\begin{eqnarray}
	Q(\Theta, \Theta^{[k]}) & = & \int_{\mbox{\boldmath $y$} \in \Upsilon}
									\log p(X,\mbox{\boldmath $y$} | \Theta) ~ p(\mbox{\boldmath $y$}|X,\Theta^{[k]}) ~~ d \mbox{\boldmath $y$} \\
							~ & = & \sum\nolimits_{\mbox{\boldmath $y$} \in \Upsilon}  \log \prod\nolimits_{i=1}^{N} w_{y_i} p(\Vec{x}_i | \theta_{y_i})  
																		\prod\nolimits_{j=1}^{N} p(y_j | \Vec{x}_j, \Theta^{[k]} ) \\
							~ & = & \sum\nolimits_{y_1=1}^{M} \sum\nolimits_{y_2=1}^{M} \cdots \sum\nolimits_{y_N=1}^{M} 
															\sum\nolimits_{i=1}^{N} \log \left[ w_{y_i} p(\Vec{x}_i | \theta_{y_i}) \right] 
															\prod\nolimits_{j=1}^{N} p(y_j | \Vec{x}_j, \Theta^{[k]} )  \label{eqn:q_fn_expanded}
\end{eqnarray}%

\noindent
It can be shown~\cite{Bilmes98} that Eqn.~(\ref{eqn:q_fn_expanded}) can be simplified to:
\begin{eqnarray}
	Q(\Theta, \Theta^{[k]}) & = & \sum\nolimits_{m=1}^{M}  \sum\nolimits_{i=1}^{N} \log[ w_m  p(\Vec{x}_i | \theta_m)] ~ p(m|\Vec{x}_i, \Theta^{[k]}) \\
							~ & = & \sum\nolimits_{m=1}^{M}  \sum\nolimits_{i=1}^{N} \log[ w_m ] ~  p(m|\Vec{x}_i, \Theta^{[k]}) +
									\sum\nolimits_{m=1}^{M}	\sum\nolimits_{i=1}^{N} \log[  p(\Vec{x}_i | \theta_m) ] ~ p(m|\Vec{x}_i, \Theta^{[k]}) ~~~ \\
							~ & = & Q_1 ~~~ + ~~~ Q_2 
\end{eqnarray}%

\noindent
Hence $Q_1$ and $Q_2$ can be maximised separately, to obtain $w_m$ and $\theta_m = \{ \Vec{\mu}_m, \Mat{\Sigma}_m \}$, respectively.
To find the expression which maximises $w_m$, we need to introduce the Lagrange multiplier~\cite{Duda01} $\psi$,
with the constraint $\sum\nolimits_m w_m = 1$, take the derivative of $Q_1$ with respect to $w_m$ and set the result to zero:
\begin{eqnarray}
	\frac{\partial Q_1}{\partial w_m} & = & 0 \\
						\therefore ~ 0  & = & \frac{\partial}{\partial w_m}
											\left\{ \sum\nolimits_{m=1}^{M} \sum\nolimits_{i=1}^{N} \log[ w_m ]  ~ p(m|\Vec{x}_i, \Theta^{[k]})
											 		+ \psi \left[ (\sum\nolimits_m w_m) -1 \right] \right\} \\
									~ & = & \sum\nolimits_{i=1}^{N} \frac{1}{w_m} ~ p(m|\Vec{x}_i, \Theta^{[k]}) + \psi \label{eqn:q1_alpha_opt}
\end{eqnarray}

\noindent
Rearranging Eqn.~(\ref{eqn:q1_alpha_opt}) to obtain a value for $\psi$:
\begin{eqnarray}
	-\psi w_m & = & \sum\nolimits_{i=1}^{N} p(m|\Vec{x}_i, \Theta^{[k]})
\end{eqnarray}

\noindent
Summing both sides over $m$ yields:
\begin{eqnarray}
	-\psi \sum\nolimits_m w_m & = & \sum\nolimits_{i=1}^{N} \sum\nolimits_m p(m|\Vec{x}_i, \Theta^{[k]}) \\
	-\psi 1 & = & \sum\nolimits_{i=1}^{N} 1 \\
	\psi & =  & -N	\label{eqn:lambda_equals_negN}
\end{eqnarray}

\noindent
By substituting Eqn.~(\ref{eqn:lambda_equals_negN}) into Eqn.~(\ref{eqn:q1_alpha_opt}) we obtain:
\begin{eqnarray}
			N & = & \sum\nolimits_{i=1}^{N} \frac{1}{w_m} ~ p(m|\Vec{x}_i, \Theta^{[k]}) \\
	\therefore ~ w_m & = & \frac{1}{N} \sum\nolimits_{i=1}^{N} p(m|\Vec{x}_i, \Theta^{[k]}) 
	\label{eqn:alpha_m_solved}
\end{eqnarray}

\noindent
To find expressions which maximise $\Vec{\mu}_m$ and $\Mat{\Sigma}_m$, let us now expand $Q_2$:
\begin{eqnarray}
	Q_2 & = &   \sum\nolimits_{m=1}^{M}	\sum\nolimits_{i=1}^{N} \log[  p(\Vec{x}_i | \theta_m) ] ~ p(m|\Vec{x}_i, \Theta^{[k]}) 	\\
	 ~  & = &   \sum\nolimits_{m=1}^{M} \sum\nolimits_{i=1}^{N} \left[
					- \frac{1}{2} \log(|\Mat{\Sigma}_m|) - \frac{1}{2} (\Vec{x}_i - \Vec{\mu}_m)^T \Mat{\Sigma}_m^{-1} (\Vec{x}_i - \Vec{\mu}_m)
					\right] p(m|\Vec{x}_i, \Theta^{[k]})  \label{eqn:q2_expanded}
\end{eqnarray}

\noindent
where $-\frac{D}{2} \log(2\pi)$ was omitted since it vanishes when taking a derivative with respect to $\Vec{\mu}_m$ or $\Mat{\Sigma}^{-1}_m$.
To find the expression which maximises $\Vec{\mu}_m$, we need to take the derivative of $Q_2$ with respect to $\Vec{\mu}_m$, and set the result to zero:
\begin{eqnarray}
	\frac{\partial Q_2}{\partial \Vec{\mu}_m} & = & 0   \\
 0 & = & \frac{\partial}{\partial \Vec{\mu}_m} \left\{ \sum\nolimits_{m=1}^{M} \sum\nolimits_{i=1}^{N} \left[
					- \frac{1}{2} \log(|\Sigma_m|) - \frac{1}{2} (\Vec{x}_i - \Vec{\mu}_m)^T \Mat{\Sigma}_m^{-1} (\Vec{x}_i - \Vec{\mu}_m)
					\right] p(m|\Vec{x}_i, \Theta^{[k]}) \right\} ~~~ ~~~ ~~~  \label{eqn:q2_mu_partone}
\end{eqnarray}%

\noindent
L\"{u}tkepohl~\cite{Lutkepohl96} states that
\mbox{$\frac{ \partial \Vec{z}^T \Mat{A} \Vec{z} }{ \partial \Vec{z} } =  (\Mat{A} + \Mat{A}^T)\Vec{z}$},
~ $(\Mat{A}^{-1})^T = (\Mat{A}^T)^{-1}$ and if $\Mat{A}$ is symmetric, then $\Mat{A} = \Mat{A}^T$.
Since $\Mat{\Sigma}_m$ is symmetric, Eqn.~(\ref{eqn:q2_mu_partone}) reduces to:
\begin{eqnarray}
	0 & = & \sum\nolimits_{i=1}^{N} - \frac{1}{2} 2 \Mat{\Sigma}_m^{-1} (\Vec{x}_i - \Vec{\mu}_m) p(m|\Vec{x}_i, \Theta^{[k]}) \\
	~ & = & \sum\nolimits_{i=1}^{N} \left[ - \Mat{\Sigma}_m^{-1} \Vec{x}_i p(m|\Vec{x}_i, \Theta^{[k]}) + \Mat{\Sigma}_m^{-1} \Vec{\mu}_m p(m|\Vec{x}_i, \Theta^{[k]})  \right] ~~~ \\
	\therefore ~ \sum\nolimits_{i=1}^{N} \Mat{\Sigma}_m^{-1} \Vec{\mu}_m p(m|\Vec{x}_i, \Theta^{[k]}) & = & \sum\nolimits_{i=1}^{N} \Mat{\Sigma}_m^{-1} \Vec{x}_i p(m|\Vec{x}_i, \Theta^{[k]}) 
\end{eqnarray}%

\noindent
Multiplying both sides by $\Mat{\Sigma}_m$ yields:
\begin{eqnarray}
	\sum\nolimits_{i=1}^{N} \Vec{\mu}_m p(m|\Vec{x}_i, \Theta^{[k]}) & = & \sum\nolimits_{i=1}^{N} \Vec{x}_i p(m|\Vec{x}_i, \Theta^{[k]})  \\
\therefore ~ \Vec{\mu}_m & = & \frac{ \sum\nolimits_{i=1}^{N} \Vec{x}_i p(m|\Vec{x}_i, \Theta^{[k]}) }
								   { \sum\nolimits_{i=1}^{N} p(m|\Vec{x}_i, \Theta^{[k]}) }
\end{eqnarray}

\noindent
L\"{u}tkepohl~\cite{Lutkepohl96} states that:  
$|\Mat{A}^{-1}| = |\Mat{A}|^{-1}$ 
and 
$\mbox{tr}(\Mat{A}\Mat{B}) = \mbox{tr}(\Mat{B}\Mat{A})$.
Since $\mbox{tr}{(\Vec{z}A\Vec{z}^T)} = \mbox{tr}(\mbox{scalar})$,
we can rewrite Eqn.~(\ref{eqn:q2_expanded}) as:
\begin{eqnarray}
	Q_2 & = &   \sum\nolimits_{m=1}^{M} \sum\nolimits_{i=1}^{N} \left[
					\frac{1}{2} \log(|\Mat{\Sigma}_m^{-1}|) - \frac{1}{2} \mbox{tr}(\Mat{\Sigma}_m^{-1} (\Vec{x}_i - \Vec{\mu}_m) (\Vec{x}_i - \Vec{\mu}_m)^T)
					\right] p(m|\Vec{x}_i, \Theta^{[k]})  \label{eqn:q2_rewritten}
\end{eqnarray}

\noindent
According to L\"{u}tkepohl~\cite{Lutkepohl96},
$\frac{\partial \log(|\Mat{A}|)}{\partial \Mat{A}} = (\Mat{A}^{T})^{-1}$
and $\frac{\partial \mbox{tr}(\Mat{B}\Mat{A})}{\partial \Mat{B}} = \Mat{A}^T$.
Moreover, we note that $\Vec{z}\Vec{z}^T$ is a symmetric matrix.
To find an expression which maximises~$\Mat{\Sigma}_m$,
we can take the derivative of Eqn.~(\ref{eqn:q2_rewritten}) with respect to $\Mat{\Sigma}_m^{-1}$
and set the result to zero:
\begin{eqnarray}
	0 & = & \frac{\partial Q_2}{\partial \Mat{\Sigma}_m^{-1}}  \\
	~ & = & \frac{\partial}{\partial  \Mat{\Sigma}_m^{-1}} \left\{
		\sum\nolimits_{m=1}^{M} \sum\nolimits_{i=1}^{N} \left[
					\frac{1}{2} \log(|\Mat{\Sigma}_m^{-1}|) - \frac{1}{2} \mbox{tr}\left(\Mat{\Sigma}_m^{-1} (\Vec{x}_i - \Vec{\mu}_m) (\Vec{x}_i - \Vec{\mu}_m)^T \right)
					\right] p(m|\Vec{x}_i, \Theta^{[k]})
	\right\} \\
	~ & = & \sum\nolimits_{i=1}^{N} \left[ \frac{1}{2} \Mat{\Sigma}_m - \frac{1}{2} (\Vec{x}_i - \Vec{\mu}_m) (\Vec{x}_i - \Vec{\mu}_m)^T \right] p(m|\Vec{x}_i, \Theta^{[k]}) \\
\end{eqnarray}%

\noindent
thus
\begin{eqnarray}
	\frac{1}{2} \Mat{\Sigma}_m \sum\nolimits_{i=1}^{N} p(m|\Vec{x}_i, \Theta^{[k]}) & = & \frac{1}{2} \sum\nolimits_{i=1}^{N} 
																		(\Vec{x}_i - \Vec{\mu}_m) (\Vec{x}_i - \Vec{\mu}_m)^T p(m|\Vec{x}_i, \Theta^{[k]}) \\
	\therefore ~ \Mat{\Sigma}_m & = & \frac{\sum\nolimits_{i=1}^{N} (\Vec{x}_i - \Vec{\mu}_m) (\Vec{x}_i - \Vec{\mu}_m)^T p(m|\Vec{x}_i, \Theta^{[k]})}
						{\sum\nolimits_{i=1}^{N} p(m|\Vec{x}_i, \Theta^{[k]})}
\end{eqnarray}

\noindent
In summary,
\begin{eqnarray}
	w_m^{[k+1]}	& = & \frac{1}{N} \sum\nolimits_{i=1}^{N} p(m|\Vec{x}_i, \Theta^{[k]})  \label{eqn:copy_of_solved_alpha}  \\  
	\Vec{\mu}_m^{[k+1]} & = & \frac{ \sum\nolimits_{i=1}^{N} \Vec{x}_i ~ p(m|\Vec{x}_i, \Theta^{[k]}) }
								   { \sum\nolimits_{i=1}^{N} p(m|\Vec{x}_i, \Theta^{[k]}) }  \label{eqn:copy_of_solved_mu} \\
	\Mat{\Sigma}_m^{[k+1]}	& = & \frac{\sum\nolimits_{i=1}^{N} (\Vec{x}_i - \Vec{\mu}_m^{[k+1]}) (\Vec{x}_i - \Vec{\mu}_m^{[k+1]})^T p(m|\Vec{x}_i, \Theta^{[k]})}
						{\sum\nolimits_{i=1}^{N} p(m|\Vec{x}_i, \Theta^{[k]})}  \label{eqn:copy_of_solved_sigma}
\end{eqnarray}

\noindent
where
\begin{equation}
	p(m | \Vec{x}_i, \Theta^{[k]}) = \frac{ p(\Vec{x}_i | \theta_{m}^{[k]}) p(m | \Theta^{[k]}) }{ \sum\nolimits_{n=1}^{M} p(\Vec{x}_i | \theta_{n}^{[k]}) p(n | \Theta^{[k]})}
\end{equation}
which can be explicitly stated as:
\begin{equation}
	p(m | \Vec{x}_i, \Theta^{[k]}) = \frac{ {\mathcal{N}}( \Vec{x}_i | \Vec{\mu}_m^{[k]},  \Mat{\Sigma}_m^{[k]} )  w_m^{[k]} }
										{ \sum\nolimits_{n=1}^{M} {\mathcal{N}}( \Vec{x}_i | \Vec{\mu}_n^{[k]}, \Mat{\Sigma}_n^{[k]} ) w_n^{[k]} }
\end{equation}

\noindent
If we let $l_{m,i} = p(m | \Vec{x}_i, \Theta^{[k]})$ and $L_m = \sum\nolimits_{i=1}^{N} l_{m,i} $, we can restate Eqns.~(\ref{eqn:copy_of_solved_alpha})
to~(\ref{eqn:copy_of_solved_sigma}) as:
\begin{eqnarray}
	w_m^{[k+1]}	& = & \frac{L_m}{N} \\  
	\Vec{\mu}_m^{[k+1]} & = & \frac{1}{L_m} \sum\nolimits_{i=1}^{N} \Vec{x}_i ~ l_{m,i} \\
	\Mat{\Sigma}_m^{[k+1]}	& = & \frac{1}{L_m} \sum\nolimits_{i=1}^{N} (\Vec{x}_i - \Vec{\mu}_m^{[k+1]}) (\Vec{x}_i - \Vec{\mu}_m^{[k+1]})^T  l_{m,i}
\end{eqnarray}

\newpage
\renewcommand{\baselinestretch}{1.06}\small\normalsize
\small
\bibliographystyle{ieee}
\bibliography{refs}

\end{document}